\documentclass[aps,pra,twocolumn,showpacs,amsmath,amssymb,floatfix]{revtex4}
\usepackage{graphicx}
\usepackage{bm}
\usepackage{color}
\def\be{\begin{equation}}
\def\ee{\end{equation}}
\def\bea{\begin{eqnarray}}
\def\eea{\end{eqnarray}}
\def\la{\langle}
\def\ra{\rangle}

\begin{document}

\title{Entanglement genesis under continuous parity measurement}
\author{Nathan S. Williams and Andrew N. Jordan}
\affiliation{Department of Physics and Astronomy, University of Rochester, Rochester, New York 14627, USA}
\date{\today}
\begin{abstract}
We examine the stochastic dynamics of entanglement for an uncoupled two-qubit system, undergoing continuous parity measurement.  Starting with a fully mixed state, the entanglement is zero for a finite amount of time, when it is suddenly created, which we refer to as entanglement genesis.  There can be further entanglement sudden death/birth events culminating in the formation of a fully entangled state.  We present numerical investigations of this effect together with statistics of the entanglement genesis time in the weak measurement limit as well as the quantum Zeno limit.  An analytic treatment of the physics is presented, aided by the derivation of a simple concurrence equation for Bell basis X-states.  The probability of entanglement border crossing and mean first passage times are calculated for the case of measurement dynamics alone.   We find that states with almost the same probability of entanglement border crossing can have very different average crossing times.  Our results provide insights on the optimization of entanglement generation by measurement.
\end{abstract}
\maketitle

Entanglement is arguably the most fascinating aspect of quantum mechanics and plays a central role in quantum information science.  The peculiar non-classical correlations which entanglement characterizes are the basis for exciting applications such as teleportation, quantum encryption, and many others \cite{NielsenChuang}.
Since the concept of entanglement appeared, there has been much research into its properties.  One aspect of this research has been how to quantify the amount of entanglement contained in a quantum state. Many different entanglement measures have been introduced with varying degrees of success \cite{Plenio2007}.  In the two qubit case, the concurrence measure of entanglement can be explicitly given in general for any mixed state \cite{wootters}, making this system ideal for further research.

In parallel with these measures, investigations into the dynamics of entanglement began and continue still.  If entanglement is to be used as a resource, then its dynamical evolution must be understood.  Of particular interest is how entanglement decays when the entangled systems are coupled to the environment.  It was found that entanglement typically decreases at an exponential rate, faster than single-qubit decoherence \cite{yu2002,yu2003,Kofman2008,Zyczk2001}.  An important step was the realization by Jak\'obczyk \cite{Jakobczyk2002} and Yu and Eberly \cite{yu2004} that entanglement can reach zero in a finite time.  This behavior was dubbed ``entanglement sudden death'' \cite{yu2004,Roszak2006,Yonac2006,Yu2006,Li2008a,Lopez2008,Ficek2006} and has received much recent attention.  The ability of entanglement to have this non-analytic behavior can be traced back to its definition: two particles are considered entangled if their combined density matrix cannot be decomposed into a weighted sum of pure, separable density matrices.  If such a decomposition can be found, then it is defined to be seperable, having zero entanglement.  While it is important to understand the disappearance of entanglement, we must also understand how it is generated in the first place.  This is the topic discussed here.  

\begin{figure}[htb!]
\begin{center}
\leavevmode 
\includegraphics[width=8cm]{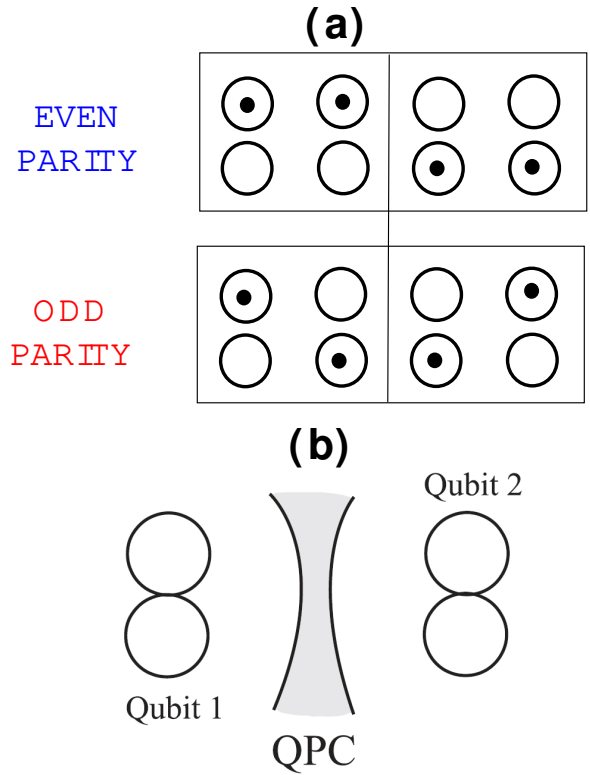} 
\caption{a) Representation of the different Parity states.  b) Diagram of a Parity Meter with two parallel double quantum dots on either side of a Quantum Point Contact (QPC)} 
\label{fig1}
\end{center}
\vspace{-5mm}
\end{figure}

Entanglement is often created via interaction-based proposals or with two-particle informational ambiguity such as parametric photon down-conversion \cite{MandelWolf} (see e.g. Refs.~\cite{Pashkin2003,Steffen2006,Samuelsson2004,Samuelsson2008} for solid-state examples). From a fundamental perspective a fascinating alternative is the case in which entanglement is generated through a measurement itself either in a solid-state \cite{ruskov2003,Trauzettel2006,Li2008,Mao2004} or quantum optics based environment \cite{Cabrillo1999,Plenio1999,Browne2003,Feng2003}.  In the simplest case we can consider a detector which measures in the two-qubit Bell basis.  If the system begins in a non-entangled state and this detector measures the system, projecting into one of its basis states, then the system becomes entangled due to the measurement.  However, rarely in actual experiments are the measurements perfectly projective, which leads us to the study of continuous measurement.  

The subject of the creation of entanglement has been examined in recent years.  Some papers investigate the idea of entanglement revivals \cite{Lopez2008,Ficek2006,Zyczk2001} where the system is initially entangled, loses its entanglement, and then regains it through a revival or ``entanglement sudden birth'' event.   While these effects are dynamically interesting, they might be better characterized as ``entanglement sudden transfer''.  This is because entanglement exists in other degrees of freedom that becomes transfered to the part of the system being examined, as pointed out in Ref.~\cite{Lopez2008}.    In contrast, the current paper looks at measurement-induced genesis of entanglement: there is no reservoir of entanglement, it is creation {\it ex nihilo} by the act of observation.  This work is closely related to quantum optics investigations of entanglement creation through a monitored decay process, see Refs.~\cite{Jakobczyk2002,Ficek2008a}.

In this paper, we focus on the {\it parity meter} (a detector which can only differentiate between qubits being aligned or anti-aligned) to accomplish entanglement generation.  It has been shown that parity meters provide a new realization of quantum computing \cite{Beenakker2004,Ionicioiu2007} and are capable of building up many different types of entangled states \cite{Trauzettel2006,Ionicioiu2008,Li2008}. 
Parity meters have also been recently been used to explore rapid two-party purification and enhanced entanglement generation using feedback control  \cite{Hill2008}.

We explore entanglement dynamics arising as a result of continuous parity measurements and more specifically analyze the entanglement genesis event.  Although our results apply to any kind of two-qubit system, for concreteness we focus on a solid state system:  a pair of quantum dot charge qubits coupled to a parity charge-sensor.  The parity in this case is defined as being even (odd) if the qubit electrons are aligned (anti-aligned) as in Fig.~1a.  There are several explicit examples of a charge-sensing parity meter, such as a quantum point contact with parallel double quantum dots on either side (see Fig.~1b) \cite{Trauzettel2006}.  See Refs.~\cite{Mao2004,Engel2005,Ruskov2006,Ionicioiu2007} for other physical implementations in both charge and spin quantum dot qubits.  The partial collapse of the system arising from the weak continuous measurement has recently been tested experimentally in a solid state environment \cite{Katz2006,Katz2008} showing that the techniques required for such a measurement are attainable with current technology and methods.

The results given in the text provide predictions about the stochastic production of entanglement from a continuous parity measurement.  This kind of entanglement generation has the counterintuitive advantage that stronger coupling both preserves and stabilizes the entangled state.  Like the Zeno effect, any tendency for the entanglement to decay from coupling to other degrees of freedom is suppressed by the act of continuous parity observation.  Our results are also useful for the purposes of generating and optimizing entanglement production.  For example, it may be easy to prepare certain initial states that minimize the time to cross the entanglement border (\ref{avgenttime}), given a fixed probability of entanglement genesis (\ref{egprob},\ref{sdprob}).

\section{Model} Under the assumption that the qubits are spatially separated and screened by the quantum point contact (QPC), any direct qubit-qubit interactions are suppressed leading to the qubit Hamiltonian $H = \left[\epsilon_1 \sigma_z^1+\epsilon_2 \sigma_z^2 + \Delta_1 \sigma_x^1+\Delta_2 \sigma_x^2 \right]/2$, where $\Delta_i$ is the tunnel coupling energy between the dots, which is set to be symmetric ($\Delta_1 = \Delta_2 = \Delta$).  For simplicity assume that the qubits are tuned such that the energy asymmetry between the levels ($\epsilon_i$) is set to zero.  This Hamiltonian gives a qubit time scale, $T_q = 2\pi / \Delta$, and has four energy eigenvectors ($\psi_1$, $\psi_2$, $\psi_3$, $\psi_4$),
\begin{equation}
\begin{gathered}
\psi_1 = \frac{|11\rangle - |00\rangle}{\sqrt{2}}, \\
\psi_2 = \frac{|10\rangle - |01\rangle}{\sqrt{2}}, \\
\psi_3 = (|11\ra + |10\ra + |01\ra + |00\ra)/2, \\
\psi_4 = (|11\ra - |10\ra - |01\ra + |00\ra)/2,  
\end{gathered}
\end{equation}
where $\vert 10 \ra$ refers to the first qubit being in the ``up'' state and the second being in the ``down'' state (etc.).  The first two solutions share the interesting property that they both have a definite parity and therefore are stable under the measurement dynamics as well as the qubit Hamiltonians.  The second two are stationary under Hamiltonian dynamics, but not under measurement dynamics, since they are not parity eigenstates.  Another interesting aspect about the first two states is that these are Bell states.  Therefore, by combining the dynamics from these two processes (measurement and Hamiltonian evolution) we end up with two stationary states that are both maximally entangled and a third possibility, where $\psi_3,\psi_4$ continually rotate among themselves and are subjected to a continuous combination of noisy measurement and Hamiltonian dynamics.  A time-averaged electrical current alone can differentiate between these three behaviors of the system at sufficiently long times \cite{Trauzettel2006}.  

The coupling between the qubits and the QPC can be modeled with the interaction Hamiltonian, 
\be
H_{int} = (\Delta \hat{E}/2)\sigma_z^1 \sigma_z^2,
\ee
where $\Delta \hat{E}$ is a charging energy operator, whose expectation value is equal to the difference in charging energy of the even and odd parity classes.  In the tunneling regime the operator $\Delta \hat{E}$ can be associated with the standard input variables $\lambda a_R^{\dagger} a_L +H.c.$, where $\lambda$ is the coupling strength and $a_{R,L}^{\dagger}$ and $a_{R,L}$ are the creation and annihilation operators associated with the right or left side of the tunnel junction.  For more details on this model of the interaction see \cite{Trauzettel2006}.

For the systems considered here the output signal of the parity meter is the electrical current, $I(t)$, through the quantum point contact.  This current is modeled following Korotkov \cite{korotkov2001} by $I(t) = \sum_k \rho_{kk} I_k + \xi(t)$, with $\rho_{kk}$ being the density matrix element associated with the state $\vert \psi_k \ra$ and $\xi(t)$ being the random white shot noise of the QPC, described with a normal distribution with spectral density, $S_0$ defined as $\langle \xi(t) \xi(0) \rangle = 2 S_0 \delta(t)$.  Each $I_i$ corresponds to the average current that the corresponding state $\vert \psi_i \ra$ would produce.  Defining the current produced from an even (odd) parity state as $I_{E,O}$ we can introduce $\bar{I} = (I_E + I_O)/2$ and $\Delta I = I_E - I_O$.  This allows the definition of dimensionless shifted currents with $I(t) - \bar{I} = x(t) \frac{\Delta I}{2}$, where $x=\pm1$ coincides with the current from an even/odd parity state.  From this model an important time scale arises, the measurement time: $T_M = 4 S_0 / \Delta I^2$.  This is the typical time needed to differentiate between the noise and the signal from the system.  With the measurement time and the qubit time a useful parameter can be defined as $K = T_q / T_M$, which can be varied by tuning parameters of the system.  $K$ can be increased through a stronger coupling between the system and detector, as well as a higher potential barrier between the quantum dots themselves.

To analyze the entanglement we follow Wootters \cite{wootters} and define concurrence as $C={\rm max}[0,\Lambda]$, where $\Lambda = \sqrt{\lambda_1} - (\sqrt{\lambda_2}+\sqrt{\lambda_3}+\sqrt{\lambda_4})$ and $\lambda_i$ corresponds to the eigenvalues of the matrix $\rho(\sigma_y \otimes \sigma_y)\rho^*(\sigma_y \otimes \sigma_y)$, ordered such that $\lambda_1 \ge \lambda_2 \ge \lambda_3 \ge \lambda_4$.  Previous works have achieved significant analytical understanding by applying concurrence to a density matrix that maintains a general X form in the computational basis ($|00\rangle$, $|11\rangle$, $|01\rangle$, and $|10\rangle$),
\begin{equation}
\rho = 
\begin{pmatrix}
\rho_{11} && 0 && 0 && \rho_{14} \\
0 && \rho_{22} && \rho_{23} && 0 \\
0 && \rho_{23}^* && \rho_{33} && 0 \\
\rho_{14}^* && 0 && 0 && \rho_{44}
\end{pmatrix},
\label{xstate}
\end{equation}
to formulate a simple expression for the concurrence, $C = 2 \max [0, \vert \rho_{23} \vert - \sqrt{\rho_{11} \rho_{44} }, \vert \rho_{14} \vert - \sqrt{ \rho_{22} \rho_{33} } ]$ \cite{Yu2006}.  Although this expression is very useful, the density matrix we presently consider is not trapped in an ``X'' state in the computational basis.  However, using the Bell Basis defined as:
\be
\begin{gathered}
|u_1\rangle = (|11\rangle - |00\rangle)/\sqrt{2}, \\  
|u_2\rangle = (|11\rangle + |00\rangle)/\sqrt{2}, \\
|u_3\rangle = (|10\rangle + |01\rangle)/\sqrt{2}, \\  
|u_4\rangle = (|10\rangle - |01\rangle)/\sqrt{2},
\label{BellBasis}
\end{gathered}
\ee
provides a similar situation.
The isolated qubit Hamiltonian expressed in this basis is given by
\be
H = 
\begin{pmatrix}
0 && 0 && 0 && 0 \\
0 && 0 && \Delta && 0 \\
0 && \Delta && 0 && 0 \\
0 && 0 && 0 && 0
\end{pmatrix}.
\label{hamiltonian}
\ee
The structure of this Hamiltonian together with the fact that the first and last Bell states share a definite parity guarentees that if the system starts in a {\it Bell basis} X-state \eqref{xstate}, then it will remain in one.  This fact again simplifies the calculation of the concurrence.  This is an important difference between previous work and our current investigation which does provide significant differences.

The eigenvalues for $\rho(\sigma_y^1 \otimes \sigma_y^2)\rho^*(\sigma_y^1 \otimes \sigma_y^2)$ corresponding to this general X-state are
\begin{widetext}
\be
\begin{gathered}
\sqrt{\lambda_{a,b}} = \frac{1}{2}\Big[\sqrt{(\rho_{11}+\rho_{44}+\rho_{14}+\rho_{14}^*)(\rho_{11}+\rho_{44}-\rho_{14}-\rho_{14}^*)} \mp \sqrt{(\rho_{11}-\rho_{44}+\rho_{14}-\rho_{14}^*)(\rho_{11}-\rho_{44}-\rho_{14}+\rho_{14}^*)}\Big]  \\
\sqrt{\lambda_{c,d}} = \frac{1}{2}\Big[\sqrt{(\rho_{22}+\rho_{33}+\rho_{23}+\rho_{23}^*)(\rho_{22}+\rho_{33}-\rho_{23}-\rho_{23}^*)} \mp \sqrt{(\rho_{22}-\rho_{33}+\rho_{23}-\rho_{23}^*)(\rho_{22}-\rho_{33}-\rho_{23}+\rho_{23}^*)}\Big].
\end{gathered}
\ee
\end{widetext}
From these eigenvalues it is clear that $\sqrt{\lambda_b} \ge \sqrt{\lambda_a}$ and $\sqrt{\lambda_d} \ge \sqrt{\lambda_c}$, making either $\sqrt{\lambda_b}$ or $\sqrt{\lambda_d}$ compete for the role of $\sqrt{\lambda_1}$.  With these two orderings we have two possible $\Lambda$ functions that compete to be the concurrence.  In spite of the lengthy expressions for the $\sqrt{\lambda}$'s, a pleasing simplicity emerges to give a concise and physically appealing result.  We expect that this result will be of interest in its own right.  We assume that the initial state is fully mixed ($\rho = \openone/4$), which means initially $\rho_{14}$ and $\rho_{23}=0$ and from the dynamics \eqref{hamiltonian} it can be seen that $\rho_{14}(t)=0$ for all $t$ and that $\rho_{23}$ will remain completely imaginary.  With this assumption one possible $\Lambda$ function is split further due to the term, $\sqrt{(\rho_{11} - \rho_{44})^2} = \vert \rho_{11} - \rho_{44} \vert$, yielding three possible values for $\Lambda$:
\be
\begin{gathered}
\Lambda_1 = 2\rho_{11}-1 \\
\Lambda_2 = 2\rho_{44}-1 \\
\Lambda_3 = \sqrt{(\rho_{22}-\rho_{33})^2+4|\rho_{23}|^2}+\rho_{22}+\rho_{33}-1.
\end{gathered}
\label{lambdas}
\ee
The correct $\Lambda_i$ to use for the concurrence 
\begin{equation}
C=\max[0,\Lambda]
\end{equation}
is the maximum of the three
\begin{equation}
\Lambda={\rm max}[\Lambda_1,\Lambda_2,\Lambda_3].
\end{equation}  
Physically these three different functions can be understood as the concurrence coming from the three different possible asymptotic outcomes of the parity meter, either one of the stable solutions with even parity ($\Lambda_1$) or odd parity ($\Lambda_2$), or the oscillatory solution ($\Lambda_3$). In the fully diagonal case we have the simple result $\Lambda = 2 \max \rho_{ii} - 1$.

\begin{figure}[htb!]
\begin{center}
\leavevmode \includegraphics[width=7.8cm]{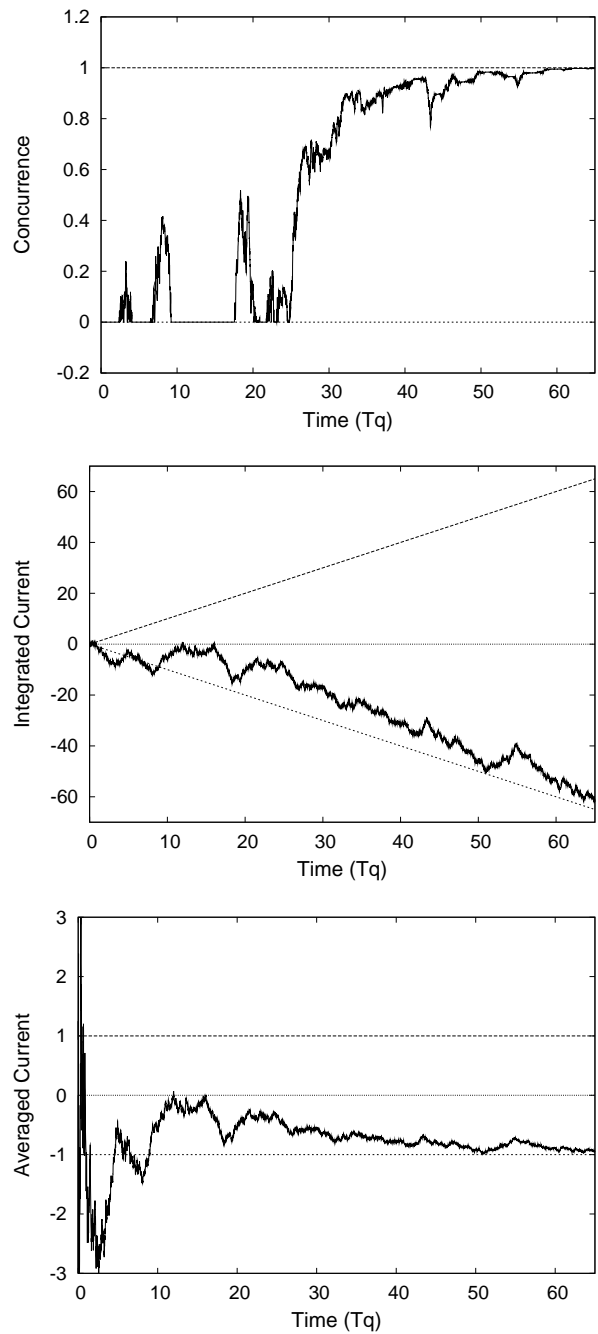} 
\caption{The top figure shows the evolution of the concurrence throughout a typical run with $K=0.3$, plotted as a function of time in units of the qubit time $T_q$.  The dashed lines represent the bounds on concurrence ($C=0$ being unentangled, $C=1$ being fully entangled).  There is always a period of quiet time punctuated by the entanglement genesis event.  Notice that there are several occurrences of entanglement sudden death/birth.  The middle figure shows the integrated detector output, with the lines representing the output from a definite purity state with no noise.  The bottom figure shows an average electrical current, with the dashed lines representing the average current from states of definite parity.}
\label{fig2}
\end{center}
\vspace{-5mm}
\end{figure}
For further examination of the measurement dynamics, we follow Korotkov, Ruskov, and Mizel \cite{Ruskov2006,korotkov2001} and write a stochastic differential equation (in It\^{o} representation) governing the evolution of the system:
\be
\begin{aligned}
\dot{\rho}_{ij} = \Big[I(t)-\sum_k \rho_{kk} I_k \Big] \Big(I_i + I_j - 2 \sum_k \rho_{kk} I_k \Big) \frac{\rho_{ij}}{S_0} \\ 
- \Big(\frac{(I_i - I_j)^2}{4S_0} + \gamma_{ij} \Big) - \frac{i}{\hbar}\Big[H,\rho \Big]_{ij},
\end{aligned}
\label{diffeq}
\ee
where again $I(t) = \sum_k \rho_{kk} I_k + \xi(t)$, with $\xi(t)$ being the random noise from the measuring device and $I_i$ corresponds to the current signal that the corresponding state $\vert u_i \ra$ would produce (in our normalized units the even states $I_E = I_1=I_2=1$ and the odd states $I_O = I_3=I_4=-1$).  The rates $\gamma_{ij}$ are the environment-induced decoherence rates, which we neglect in order to focus on the fundamental physics.  Due to the non-analytic nature of the concurrence we cannot create a single differential equation to govern its evolution, so we will initially concentrate on numerical analysis.  

\begin{figure}[tbh!]
\begin{center}
\leavevmode \includegraphics[width=8cm]{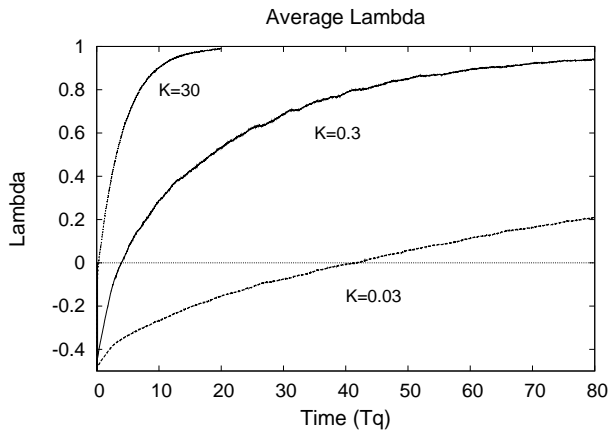} 
\caption{Average $\Lambda$ function with $K = 0.03,0.3$, and $30$, averaged over $1000$ runs.  It is useful to consider $K = 0.03,0.3,30$ to characterize when the case of a weakly coupled system ($K=0.03$), a moderate coupling ($K=0.3$) and a system that is in the strong coupling regime ($K=30$).  The entanglement genesis time is when $\Lambda$ crosses 0.} 
\label{fig3}
\end{center}
\vspace{-5mm}
\end{figure}
\section{Numerical Analysis}
Starting from an initial fully mixed state, $\rho_{\rm in} = \openone /4$, we can allow the system to evolve governed by \eqref{diffeq}.  A typical run is displayed in Fig. 2.  We can see several interesting features in these runs.  First, the system starts out unentangled due to the initial fully mixed state and then after some amount of time experiences an entanglement genesis event.  Understanding this time scale is a primary goal of this investigation.  Secondly, we see that entanglement sudden death behavior, first examined in optical systems \cite{Jakobczyk2002,yu2004}, is a common occurrence in this system.  This sudden death behavior can be understood as the system initially being projected into one of the three outcomes from the parity meter, and then due to noise the system will begin to drift back toward a different outcome causing the system to lose its entanglement for a period of time. In addition to the sudden death, we also see the recovery of entanglement through sudden birth transitions numerous times during the time evolution. The entanglement is always regained eventually since each asymptotic solution to the dynamics is a maximally entangled state.

Another quantity that we can examine with these numerical simulations is the average $\Lambda$ function, seen in Fig.~3.  We can again see a quiet period before entanglement is generated (when $\Lambda < 0$).  This {\it entanglement genesis time} is dependent on the coupling strength of the meter.  The average concurrence asymptoticly approaches 1 in an exponential fashion.  Of interest is the spread around this average genesis time.  Running 10,000 realizations of the quantum measurement and plotting a frequency histogram (see Fig. 4) we find the following:  
The distribution is zero up until some finite point, indicating that there is {\it always} a quiet period before the entanglement is generated.  This can be understood as the system having to traverse the Hilbert space from its originally fully mixed state and cross the entanglement border to the closest state which is entangled.  Using the trace distance \cite{NielsenChuang}, defined as $d = \frac{1}{2} {\rm Tr} \vert \rho - \sigma \vert$ where $\rho$ and $\sigma$ are two states, the closest entangled state that has a non-zero trace distance of $d=1/16$ is
\begin{equation}
\sigma = 
\begin{pmatrix}
1/4 && 0 && 0 && 0 \\
0 && 1/4 && i /4 && 0 \\
0 && -i /4 && 1/4 && 0 \\
0 && 0 && 0 && 1/4
\end{pmatrix}.
\end{equation}
This state is on the boundary, $\Lambda = 0$, of entanglement and it should be noted that this state can be achieved with just the qubit Hamiltonians.  However, without the measurement dynamics, this state will never cross the border and generate finite  entanglement.

\begin{figure}[t]
\begin{center}
\leavevmode \includegraphics[width=8cm]{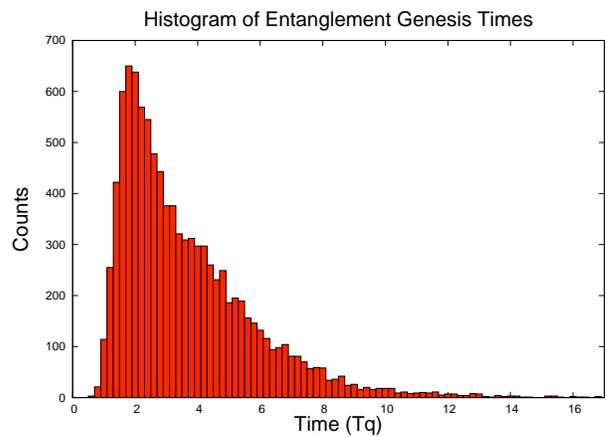} 
\caption{Histogram showing the frequency of particular entanglement genesis times from 10,000 runs, with $K=0.3$.  Each bar represents 0.2 $T_q$.  Notice the gap in entanglement times at the beginning as well as the long tail where the entanglement genesis time can be larger than $10T_q$ due to the competing $\Lambda$ functions.} 
\label{fig4}
\end{center}
\vspace{-5mm}
\end{figure}

One further numerical result is obtained in the interesting regime where the coupling strength between the measurement device and the system is very strong ($T_q / T_M \gg 1$), rapidly projecting the state.  We expect to see quantum jump behavior similar to the Quantum Zeno Effect (see Ref. \cite{Braginsky}).  Studying the numerical results in Fig.~5, this type of sudden shift in the system can be seen as well as other interesting behavior.
\begin{figure}[htb!]
\begin{center}
\leavevmode \includegraphics[width=8cm]{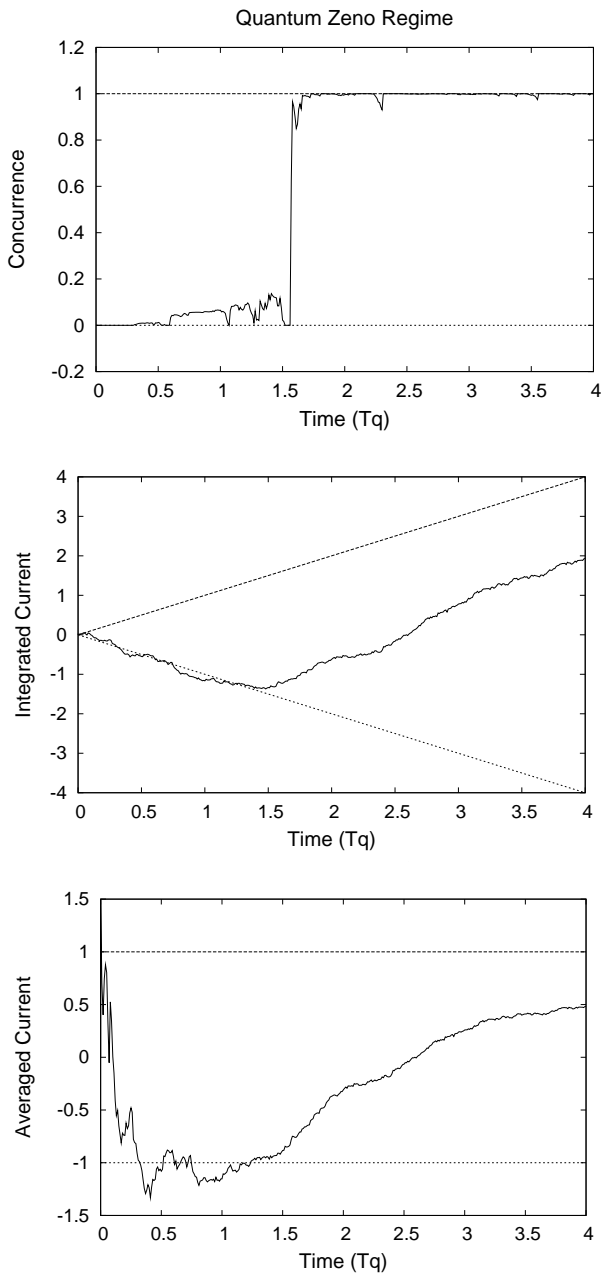} 
\caption{These figures demonstrate a run with $K = 30$ where we are in the Quantum Zeno Regime.  The top figure shows the evolution of concurrence.  The middle plot again shows the integrated output and the bottom plot shows the averaged output.  The concurrence rapidly leaps to $C=1$ under a projection into the even sub-space.} 
\label{fig5}
\end{center}
\vspace{-5mm}
\end{figure}
Very soon after the parity meter is turned on, $\Lambda$ leaps from its initial value $-1/2$ to zero, corresponding to the fully mixed state being projected quickly into the odd parity subspace ($I=-1$).  Since the system started out in a fully mixed state, this parity subspace has both states in the subspace almost equally occupied, causing the $\Lambda$ function to be around $\Lambda=0$.  Secondly, after this initial leap to zero the concurrence has a period of bouncing around near zero.  This can be understood as the qubit Hamiltonian rotating a small probability amplitude between $\vert u_2 \rangle$ and $\vert u_3 \rangle$, and then the measurement collapses the system back into the odd subspace.  A final quantum leap is seen when the small rotation to the even parity subspace causes the system to collapse toward this new outcome, resulting in the jump in concurrence and leaving the system in the oscillatory solution with $C = 1$.  

\section{Analytical Approach}
In order to have a better understanding of this physics we can take a simplified analytic approach.  Let us assume that we have projective measurements giving either ``even'' or ``odd'' at each step, with unitary Hamiltonian evolution occurring between them.  This will allow us to examine the average concurrence at each step and derive an expression for the concurrence after an arbitrary number of measurement steps.  Starting with a fully mixed state $\rho = \openone/4$, we have an initial concurrence, $C=0$.  The probability of measuring an even parity result is $\rho_{11}+\rho_{22}$ and the probability of an odd parity result is $\rho_{33}+\rho_{44}$.  Since we are now dealing with projective measurements, each measurement will fully project the system into its respective parity subspace.  The first measurement will result in
\begin{figure}[t]
\begin{center}
\leavevmode \includegraphics[width=8cm]{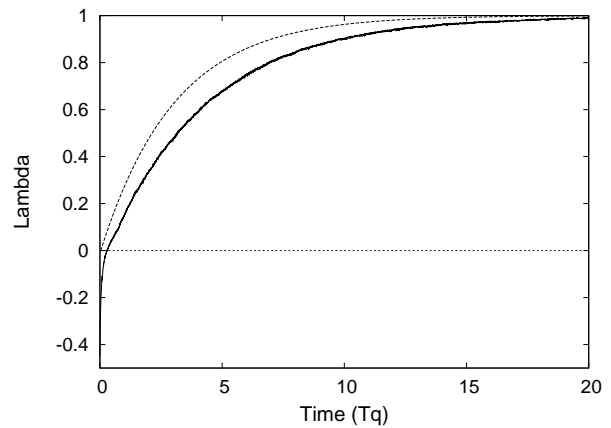}
\caption{The ensemble average of $\Lambda$ is plotted versus the qubit time $T_q$.
The solid line represents the numerical results from 10,000 runs with $K=30$.  The dotted curve is the simplified projective analytic result Eq.~\eqref{projectiveanalytic} with $\delta = \pi / K$.  Both curves start out at $\Lambda = -0.5$, however the projective model jumps to zero after one measurement time ($T_q/30$).} 
\label{fig6}
\end{center}
\vspace{-5mm}
\end{figure}
\begin{equation}
\rho_E = (1/2)\, {\rm diag}(1, 1, 0, 0),
\end{equation}
with probability $1/2$ corresponding to the first result returning ``even'' or
\begin{equation}
\rho_O = (1/2)\, {\rm diag}(0, 0, 1, 1),
\end{equation}
with probability $1/2$ corresponding to the first result returning ``odd''.  Each of these results has $\Lambda = 0$, so the average concurrence is zero for the first measurement $(n=1)$.  Now the qubit Hamiltonian dynamics comes into play and mixes the parity.  Assuming a small time step $\delta_t$ such that $\delta = \delta_t \Delta/2 \ll 1$ and that the first measurement returned an even result, we can write the evolved density matrix as
\begin{equation}
\rho'_E = \frac{1}{2}
\begin{pmatrix}
1 && 0 && 0 && 0 \\
0 && \cos^2\delta && i \cos\delta\sin\delta && 0 \\
0 && -i \cos\delta\sin\delta && \sin^2\delta && 0 \\
0 && 0 && 0 && 0
\end{pmatrix}.
\end{equation}

From here on we can just consider this case because if the first result were odd, the 
analysis is a mirror image of that below, and gives the same average.  A second projective measurement will again collapse the system into a parity subspace, where there is a probability $p_O = (1/2) \sin^2\delta$ that the result be odd, preparing the Bell state $|u_3\ra$ with concurrence $C_O=1$.  However, for $\delta \ll 1$ this probability is small, and most of the time the system will be projected back into the even subspace with probability $p_E = (1+ \cos^2\delta)/2$, preparing the state
\be
\rho''_E = \frac{1}{1+\cos^2\delta} \begin{pmatrix}
1 && 0 && 0 && 0 \\
0 && \cos^2\delta &&0 && 0 \\
0 && 0 && 0 && 0 \\
0 && 0 && 0 && 0
\end{pmatrix},
\ee
which has a concurrence of $C_E = 2 \max_j \rho''_{E,{jj}} -1 = (1-\cos^2\delta)/(1+\cos^2\delta)$.
This situation is a two-qubit analog to the Zeno effect, and similar to the discussion in Ref.~\cite{Ruskov2006}.  Once there is a result that differs from the earlier outcomes of the parity measurement, the system is then always trapped in an oscillatory combination of $|u_{2,3}\ra$, from then on contributing $C=1$.  Therefore the probability of having a sequence of the same results is the important quantity to follow in the evolution, as well as the value of the concurrence associated with those states in the sequence.  The average concurrence $\la C\ra$ after the second measurement $(n=2)$ is given by $\la C \ra = p_E C_E + p_O C_O = 1 - \cos^2 \delta$.  Generalizing this approach, we find that the average concurrence as a function of the number of measurement steps $n$ is 
\be
\la C \ra(n)=1-\cos^{2(n-1)}\delta.
\label{projectiveanalytic}
\ee  
This solution can be compared to the continuous Zeno model by considering the numerical average of $\Lambda$ over many realizations with $K=30$ by choosing $\delta = \pi / K$ and comparing it with the result plotted in Fig.~6, with reasonable agreement.  We note that the projective model rises to a non-negative concurrence after the first measurement, while the continuous measurement model takes longer to create entanglement initially, and consequently lags behind the projective model.

We next consider the case when the measurements are weak.  Unfortunately, following the same calculations as above for weak continuous measurements are rather cumbersome.  To make further analytical progress, we will adopt a strategy similar to Korotkov and one of the authors in Ref.\cite{Korotkov2006} to investigate entanglement border crossing.  We proceed by ignoring the unitary evolution from the qubit Hamiltonian dynamics and concentrate on the measurement dynamics alone.  We consider an arbitrary diagonal state in the Bell basis
\begin{equation}
\rho = 
\begin{pmatrix}
\rho_{11} && 0 && 0 && 0 \\
0 && \rho_{22} && 0 && 0 \\
0 && 0 && \rho_{33} && 0 \\
0 && 0 && 0 && \rho_{44}
\end{pmatrix}.
\end{equation}
A weak continuous measurement of the current yields new information about the system.  As a result of this information, the density matrix of the system should be updated.  Since we are neglecting the qubit Hamiltonian we can use the classical Bayes rule to update the density matrix.  This is done by interpreting the combination $p_E = \rho_{11} + \rho_{22}$ as the probability to be in the even subspace, and the combination $p_O = \rho_{33} + \rho_{44}$ as the probability to be in the odd subspace.  Given the observation of an electrical current $I$, these probabilities are updated using the conditional probabilities of finding the current $I$, given a definate parity, $P_{E,O}(I)$, 
via Bayes formula, $p_E' = P_E(I) p_E / \left[ P_E(I) p_E + P_O(I) p_O \right]$, and the same for $p_O'$.  We take $P_{E,O}(I)$ to be Gaussian distributions given by
\be
P_{E,O} = \sqrt{t/\pi S_0} \exp \Big[ -(I - I_{E,O})^2t / S_0 \Big],
\label{condprobs}
\ee
where $I_{E,O}$ is the average current from an even (odd) parity state, and introduce the rescaled measurement result 
\be
\gamma(t) = (I - I_0) (t \Delta I / S_0),
\ee
to find that the updated density matrix conditioned on the measurement result $\gamma$ is given by
\begin{equation}
\rho' = \frac{1}{\mathbb{N}}
\begin{pmatrix}
\rho_{11}{\rm e}^\gamma && 0 && 0 && 0 \\
0 && \rho_{22}{\rm e}^\gamma && 0 && 0 \\
0 && 0 && \rho_{33}{\rm e}^{-\gamma} && 0 \\
0 && 0 && 0 && \rho_{44}{\rm e}^{-\gamma}
\end{pmatrix},
\label{updated}
\end{equation}
with $\mathbb{N}=(\rho_{11}+\rho_{22}){\rm e}^\gamma+(\rho_{33}+\rho_{44}){\rm e}^{-\gamma}$.  Notice that as $\gamma \rightarrow \pm \infty$ the system collapses into the even (odd) subspace as expected.  From \eqref{lambdas}, $\Lambda$ is 
\begin{equation}
\Lambda = \frac{2}{\mathbb{N}}{\rm max}[\rho_{11}{\rm e}^\gamma,\rho_{22}{\rm e}^\gamma,\rho_{33}{\rm e}^{-\gamma},\rho_{44}{\rm e}^{-\gamma}]-1.
\label{nounilam}
\end{equation}
Entanglement border crossing happens whenever this $\Lambda$ expression crosses zero, which can correspond either to sudden death of entanglement or entanglement genesis depending on the direction.  In order to explore these phenomena we need to determine what value of $\gamma$ will correspond to $\Lambda=0$.  Setting \eqref{nounilam} equal to zero we can solve for $\gamma$, producing two solutions, $\gamma = r_1$ or $r_2$,
\begin{equation}
\begin{gathered}
r_1 = \frac{1}{2}\ln\Big[\frac{\rho_{33}+\rho_{44}}{2\max [\rho_{11},\rho_{22}]-\rho_{11}-\rho_{22}}\Big], \\
r_2 = -\frac{1}{2}\ln\Big[\frac{\rho_{11}+\rho_{22}}{2\max [\rho_{33},\rho_{44}] - \rho_{33} - \rho_{44}}\Big].
\label{rs}
\end{gathered}
\ee
\begin{figure}[tbh!]
\begin{center}
\leavevmode \includegraphics[width=9cm]{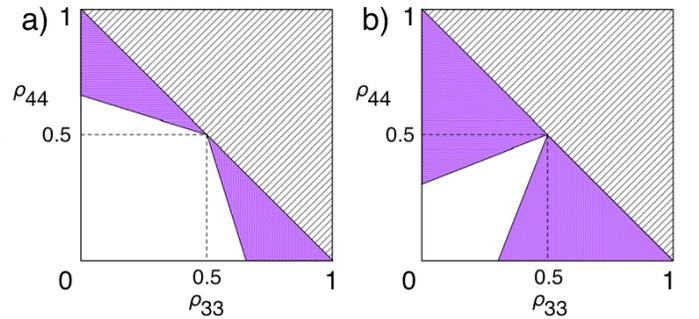}
\caption{Illustration of entanglement border crossing, conditioned on the measurement result $\gamma$.  Initially the states in the lower quadrant (marked by dashed lines) are unentangled, while states outside the lower quadrant (but still in the physical lower triangle) are entangled. (a)  $\gamma > 0$:  Some initially entangled states experience entanglement sudden death.  (b) $\gamma < 0$:  Some unentangled states experience entanglement genesis.  Entangled states are shaded in purple, and the border is given by straight lines originating from the center of the square.} 
\label{fig7}
\end{center}
\vspace{-5mm}
\end{figure}
These two solutions correspond to the possibility that starting from an unentangled state, one can become entangled via pure measurement dynamics by collapsing to either the even or the odd parity subspace.
From Eq.~(\ref{rs}) we can see that if $\rho_{11} = \rho_{22}$, or $\rho_{33} = \rho_{44}$ we will have an infinite $r_{1(2)}$, which makes physical sense.  Considering the case $\rho_{11} = \rho_{22}$, if the system completely collapses into the even parity subspace (which would correspond to $\gamma \rightarrow + \infty$) this results in the density matrix elements $\rho'_{11}=\rho'_{22}=1/2$ corresponding to an equal mixture of the first two Bell States with $\Lambda = 0$.  Therefore if one starts with $\rho_{11} = \rho_{22}$, then the even route to entanglement is blocked, leaving only the odd route.

We now analyze the measurement dynamics and find the probabilities and mean time needed to cross the entanglement border, $\Lambda=0$.  We begin by noting that the conditional probability distributions (\ref{condprobs}) are solutions to the two diffusion equations  
\be
\partial_{t} P_i = - v_i \partial_{\gamma} P_i + D \partial_{\gamma}^2 P_i, \qquad i=E,O,
\label{diffusioneq}
\ee
where $v_{E,O} = \pm 1/T_M$ correspond to the different drift velocities depending on the parity, and $D = 1/(2 T_M)$.
The probability distribution of the measured current, $P(I) = (\rho_{11} + \rho_{22}) P_E(I) + (\rho_{33} + \rho_{44}) P_O(I)$, does not involve any off-diagonal density matrix elements, and therefore the calculation is identical to a classical system with probabilities $p_{E,O}$ of being in state ``even'' or ``odd''.  We may therefore solve the problem for each case separately, and then weight the solutions with the probabilities $p_E, p_O$ of occupying a given parity.

We wish to calculate the probability that the random variable $\gamma(t)$ crosses the specific values $r_{1,2}$ by solving the diffusion equations using a Green function approach.  This approach places absorbing boundary conditions at $\gamma = r_{1,2}$ in order to find the first passage time statistics.  The details of this calculation are in the Appendix.  Several important results emerge from the analysis.  As described previously, we neglect single-qubit Hamiltonian dynamics $(\Delta=0)$ and assume that the initial state is diagonal in the Bell basis with $\rho_{11}=\rho_{22}\ne 0$ and $\rho_{33} \ne \rho_{44}$. 
The condition $\rho_{11}=\rho_{22}\ne 0$ together with the normalization of the state reduces the number of independent state variables to 2, which we take to be $\rho_{33}, \rho_{44}$.  States within the lower quadrant of the $\rho_{33}-\rho_{44}$ plane are seperable, while physical states outside the quadrant are entangled.  The entanglement border is drawn as a dashed line in Fig.~7.  As the measurement progresses, the updated state (\ref{updated}) can cross the entanglement border, depending on the value of $\gamma$.  In Fig.~7 we indicate the initial states that cross the entanglement border for (a) $\gamma > 0$ (where some states experience entanglement sudden death), and (b) $\gamma  < 0$ (where some states experience entanglement genesis).  The entangled states are shaded in purple, and the border is given by the lines 
$\rho_{44} = a \rho_{33} + b$, where $a = (1 \pm e^{2 \gamma})/(1 \mp e^{2\gamma})$, $b= e^{2\gamma}/(1 \mp e^{2\gamma})$ for the cases $\rho_{44} \lessgtr \rho_{33}$.
\begin{figure}[tbh!]
\begin{center}
\leavevmode \includegraphics[width=8cm]{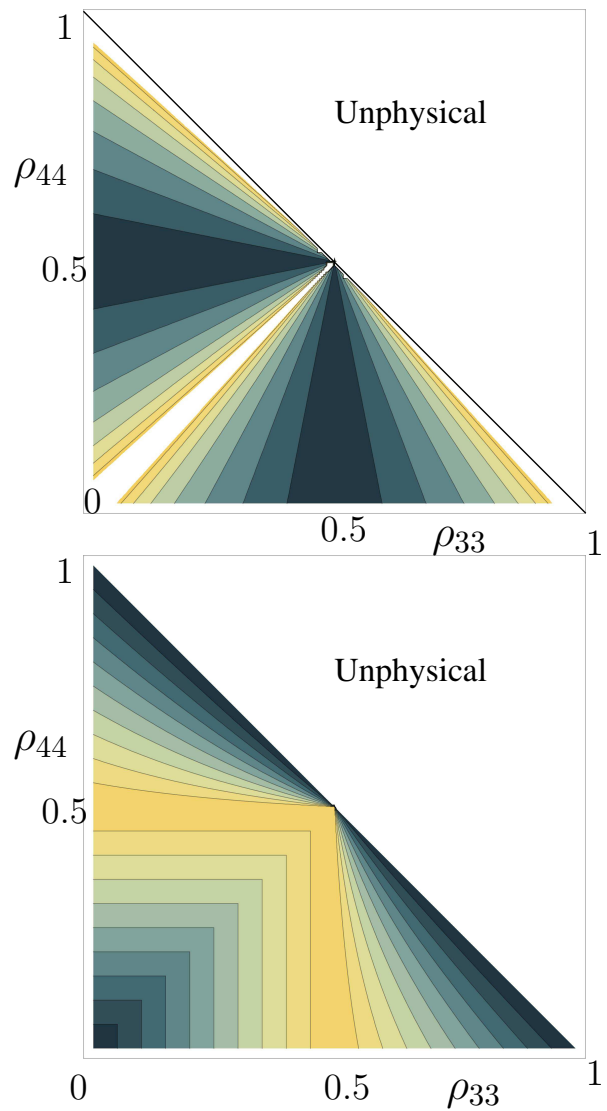}
\caption{Top: The average entanglement border crossing time $T_C$ is plotted in units of the measurement time, $T_M$, as a function of $\rho_{33}$ and $\rho_{44}$ for states where $\rho_{11} = \rho_{22} \ne 0$ and $\rho_{33} \ne \rho_{44}$.  The contours are set at time increments of 0.2 with dark blue (0) to yellow (1.4) (white exceeds 1.4).
Bottom: Probability of an entanglement border crossing ($P_{EG}$ or $P_{SD}$ depending on the initial state) as a function of $\rho_{33}$ and $\rho_{44}$.  Notice the change in behavior between EG and SD across the lower quadrant boundary.  The contours are set at every 0.1 probability increment with dark blue to yellow representing 0 to 1.} 
\label{fig8}
\end{center}
\vspace{-5mm}
\end{figure}

We now present our results for the specific case discussed above.  The probability to begin in an unentangled state (in the lower quadrant of the $\rho_{33}-\rho_{44}$ plane) and to have an entanglement genesis (EG) event at any point in time is given by 
\be
P_{EG} = 2 \max [\rho_{33},\rho_{44}]. 
\label{egprob}
\ee
If we start from an entangled state (outside the lower quadrant) we find the probability for a sudden death (SD) event at any point in time to be 
\be
P_{SD} = P_{EG} \frac{1-\rho_{33} -\rho_{44}}{|\rho_{33} - \rho_{44}|}.  
\label{sdprob}
\ee
It is remarkable that the results naturally break into two parts, depending on the direction of the border crossing, and further are independent of all details except the initial density matrix elements. The reason for the difference between $P_{EG}$ versus $P_{SD}$ can be intuitively understood by considering the extremes.  The entangled states close to the line $\rho_{33} + \rho_{44}=1$ can only experience sudden death if the system collapses toward the even parity state.  However, $\rho_{33} + \rho_{44}$ is the probability to have odd parity, so these states are unlikely to experience sudden death.  The other extreme corresponds to unentangled states close to the origin.  These states can only experience entanglement genesis if the system collapses toward the odd parity state.  However, being close to the origin means that the probability to be in the odd subspace is small, therefore entanglement genesis is unlikely for these states.

The average time, $T_C$, it takes the system to cross the entanglement border conditioned on the fact that it will cross can be determined from the first passage time probability (calculated in the Appendix)
\be
P_{fpt|C} = \frac{|r_2|}{\sqrt{2 \pi \tau^3}} \exp \Big( - \frac{(|r_2| - \tau)^2}{2 \tau} \Big),
\ee
and is found to be 
\be
T_C = T_M \vert r_2 \vert = \frac{T_M}{2} \left\vert \ln \left[\frac{1-\rho_{33}+\rho_{44}}{\vert \rho_{33} - \rho_{44} \vert}\right] \right\vert,
\label{avgenttime}
\ee
from \eqref{crossprob} and for {\it either} sudden death or entanglement genesis.  Plots of $T_C$ and $P_{EG}$, $P_{SD}$ can be seen in Fig.~8.
The above equations relate to the system being collapsed into the odd subspace to create or lose entanglement, but by symmetry if the assumptions involved are reversed: $\rho_{33}=\rho_{44} \ne 0$, $\rho_{11} \ne \rho_{22}$, then $\rho_{11, 22}$ would exchange roles with $\rho_{33, 44}$ in the above equations, and the system would collapse into the even parity subspace to create or lose entanglement.

For clarity and further insight, we now consider the specific cases of: 

i) an almost fully mixed state with 
\be
\rho_{11} = \rho_{22} = 1/4 + \epsilon \text{, } \rho_{33}=1/4-3\epsilon \text{, and } \rho_{44}=1/4 + \epsilon,
\ee
 ii) a near-Bell state with 
\be
\rho_{11}=\rho_{22}=\epsilon \text{, } \rho_{33} = 0 \text{, } \rho_{44} = 1-2\epsilon,
\ee
and a state close to the boundary on either side (non-entangled) iiia) with
\be
\rho_{11} = \rho_{22} = 1/4 \text{, } \rho_{33} = \epsilon \text{, } \rho_{44} = 1/2 - \epsilon,
\ee 
and (entangled) iiib) with 
\be
\rho_{11} = \rho_{22} = 1/4-\epsilon \text{, } \rho_{33} = \epsilon \text{, } \rho_{44} = 1/2 + \epsilon.
\ee
For these states, we find that as $\epsilon \to 0$:
\begin{itemize}
\item[i)]
\begin{itemize}
\item[ ] Almost fully mixed:
\item[ ] $P_{EG} = 1/2 +2\epsilon \to 1/2$,
\item[ ] $T_C = \frac{T_M}{2} \left\vert \ln \left[ \frac{1/2+2 \epsilon}{4 \epsilon} \right] \right\vert \to \infty$.
\end{itemize}
\item[ii)]
\begin{itemize}
\item[ ] Near-Bell state:
\item[ ] $P_{SD} = 4 \epsilon \to 0$,
\item[ ] $T_C = \frac{T_M}{2} \left\vert \ln \left[ \frac{2\epsilon}{1-2\epsilon}\right] \right\vert \to \infty$.
\end{itemize}
\item[iiia)]
\begin{itemize}
\item[ ] Non-entangled state near border:
\item[ ] $P_{EG} = 1 - 2\epsilon \to 1$,
\item[ ] $T_C = \frac{T_M}{2} \left\vert \ln \left[ \frac{1/2}{1/2 -2\epsilon}\right]\right\vert\to 0$.
\end{itemize}
\item[iiib)]
\begin{itemize}
\item[ ] Entangled state near border:
\item[ ] $P_{SD} = (1/2 - 2 \epsilon) \left( 1 + \frac{1/2 + 2 \epsilon}{1/2} \right) \to 1$,
\item[ ] $T_C = \frac{T_M}{2} \left\vert \ln \left[ \frac{1/2 - 2 \epsilon}{1/2} \right] \right\vert \to 0$.
\end{itemize}
\end{itemize}
It is interesting that in case (i), the state will become entangled with $1/2$ probability, but the time required for this to happen diverges logarithmically. 
The reason for this can be understood from Fig.~7.  For the near-fully mixed state, half of the time the system collapses into the odd parity state, so $\gamma \rightarrow -\infty$.  In Fig.~7(b) the entanglement border is pinching in on the state located near $(1/4, 1/4)$.  The entanglement border approaches the state in an exponentially decaying way, and that is why the crossing time is logarithmically divergent.  Another surprising result from these calculations is that some states take a long time to cross the entanglement border, but have the same crossing probability as states that take a very short time.  For example, if $\rho_{11} = \rho_{22} = .25$, $\rho_{33} = .49$, $\rho_{44} = .01$, we find that $T_C = (T_M /2) \ln [ 50/48 ] \approx 0.02 T_M$ and $P_C = .98$.  However, with a different state, $\rho_{11}=\rho_{22}=.02$, $\rho_{33}=.49$, $\rho_{44}=.47$, we find the same $P_C$ but the crossing time becomes $T_C = (T_M / 2) \ln [ 2 ] \approx 0.35 T_M$.  Two states with the same probability of crossing have an order of magnitude difference in their average crossing times.



\section{Conclusion}
In this paper the entanglement genesis and sudden death behavior of a two-qubit system coupled to a parity meter has been modeled and investigated.  We found from numerical simulation that both effects are generic.  We developed a simplified analytical approach to look at the behavior of the concurrence as well as the typical time scales involved in the entanglement generation.  We found the probabilities of entanglement genesis and sudden death (\ref{egprob},\ref{sdprob}), as well as the average time of entanglement border crossing \eqref{avgenttime}.  When starting from a fully mixed state there will always be some duration of time before entanglement is generated due to the distance between the initial and nearest entangled state in the Hilbert Space.  This entanglement genesis time can vary greatly for different states, even though the probability of having such an event remains comparable.  The detailed understanding of entanglement creation in these systems sheds light on the fundamental nature of continuous entanglement dynamics, as well as providing important aids in the development of methods to control and optimize entanglement generation.  A fascinating topic for further research is the use of feedback to steer the entanglement production.  A primary goal of this research would be to prevent entanglement sudden death, or to shorten the entanglement genesis time.

\appendix*
\section{}
The purpose of this Appendix is to explain the Green function approach to calculating the probabilities of entanglement border crossing, and the associated first crossing time statistics.  We begin by rescaling time in units of the measurement time, $t = T_M \tau$, so the conditional probabilities \eqref{condprobs} of a measurement outcome $\gamma$ associated with a given parity subspace are given by
\be
P_{E,O}(\gamma)=\sqrt{\frac{1}{2 \pi \tau}}{\rm exp}\Big(-\frac{(\gamma \mp \tau)^2}{2\tau}\Big).
\ee
These distributions are solutions to two random walks with drift velocities $v_{E,O}=\pm 1$ and a diffusion constant of $D=1/2$.  These random walks are described by the Fokker-Planck equations
\be
\partial_\tau P_i = -v_i \partial_\gamma P_i + D \partial_\gamma^2 P_i, \quad i = E,O.
\label{fokkerplanck}
\ee
If we want to look at the first passage statistics relating to $\Lambda$ crossing zero then we must first find the Green functions, $G_i$ which satisfy \eqref{fokkerplanck}, with the initial condition that $\gamma=0$.  Following the argument given in the body of the paper, the solutions of the Green functions $G_i(\gamma, \tau)$ for a definite parity $i=E,O$ are weighted by the even/odd parity probabilities $p_E=\rho_{11}+\rho_{22}$ and $p_O=\rho_{33}+\rho_{44}$ respectively in the final results.  We place absorbing boundary conditions at $\gamma=r_1$ and $r_2$ to account for the fact that we are only interested in the first passage time (for more details see Ref.~\cite{Redner}).  This gives us the condition that $G_i(\gamma =r_{1,2},\tau)=0$.  Following the discussion in the text, we now focus on the case $\rho_{11} = \rho_{22} \ne 0$ and $\rho_{33} \ne \rho_{44}$ in order to simplify the analysis, because $r_1 \rightarrow \infty$ giving only one finite absorbing boundary condition at $\gamma = r_2$.  We recall that gaussian functions are solutions to differential equations of this type so we can find the Green functions by examination:
\be
\begin{gathered}
G_i(\gamma,\tau)= \frac{1}{\sqrt{4 \pi D \tau}} {\rm exp}\Big(-\frac{(\gamma-v_i \tau)^2}{4D\tau}\Big) \\
\times \Big[1-{\rm exp}\Big(-\frac{r_2^2 - r_2 \gamma}{D \tau}\Big)\Big].
\end{gathered}
\ee
The above equation satisfies the differential equation of \eqref{fokkerplanck} as well as the necessary boundary and initial conditions.  We now focus on the entanglement genesis case and take $r_2$ to be negative.  The function $P_{<0}^i(\tau) = \int_{r_2}^\infty G_i(\gamma, \tau) d\gamma$ is defined as the total probability that $\Lambda$ is less than zero at time $\tau$ for a given initial state.  In the sudden death case, $r_2$ is positive so the system is initially entangled.  Here, we define the function $P_{>0}^i(\tau) = \int_{-\infty}^{r_2} G_i(\gamma, \tau) d\gamma$ to be the probability for $\Lambda$ remaining larger than zero.  For definiteness, we now focus on the entanglement genesis (EG) case (the sudden death (SD) case is directly analogous).
Since there is one finite absorbing boundary, the only place to lose probability is at $\gamma = r_2$.  Therefore, the first-passage time probability distribution $P_{fpt}^i(\tau)$ is given by
\be
\begin{aligned}
P_{fpt}^i (\tau) &= -\partial_\tau P_{<0}^i(\tau), \\
&= - \partial_\tau \int_{r_2}^\infty G_i(\gamma, \tau) d\gamma, \\
&= - \int_{r_2}^\infty \partial_\tau G_i d\gamma.
\end{aligned}
\ee
From \eqref{fokkerplanck} we have that $\partial_\tau G_i = -v_i \partial_\gamma G_i + D \partial_\gamma^2 G_i$ so we can rewrite the above equation as
\be
\begin{aligned}
P_{fpt}^i(\tau) &= - \int_{r_2}^\infty \partial_\gamma \Big( -v_i G_i + D \partial_\gamma G_i \Big) d\gamma, \\
&= - \Big[-v_iG_i + D \partial_\gamma G_i \Big] \Big|_{r_2}^\infty, \\ 
&= D \partial_\gamma G_i |_{r_2}, \\
&= \frac{ |r_2|}{\sqrt{4 D \pi \tau^3}}\exp \Big(-\frac{(r_2 - v_i \tau)^2}{4D \tau}\Big),
\end{aligned}
\label{fptprob}
\ee
where we have used the boundary condition $G_i(\gamma = r_2) =0$.  The SD case is accounted for with the absolute value of $r_2$.  Integrating this result over all positive time will give us the probability that the $\Lambda = 0$ point is crossed eventually, 
\be
P_{cross}^i = \exp\left(\frac{ r_2 v_i - |r_2 v_i|}{2D}\right).
\label{crossingprob}
\ee
The EG case corresponds to $r_2<0$, and we can therefore find $P_{EG}^i$ (the probability of entanglement genesis given parity state $i=E,O$) is given by
\be
P_{EG}^i = 
\begin{cases}
e^{2 r_2}, & \text{for $i = E$,} \\
1, & \text{for $i = O$},
\end{cases}
\label{crossingprob2}
\ee
while the SD case corresponds to $r_2>0$, and we can therefore find $P_{SD}^i$ (the probability of entanglement genesis given parity state $i=E,O$) is given by
\be
P_{SD}^i = 
\begin{cases}
1, & \text{for $i = E$,} \\
e^{-2 r_2}, & \text{for $i = O$}.
\end{cases}
\label{crossingprob3}
\ee
The probability of 1 makes physical sense because for the (odd,even) parity states, the (negative,positive) drift velocity $v_2 = \mp 1$ will eventually send $\gamma$ past $r_2$.  The exponentially suppressed probabilities correspond to the noise fighting against the drift term.  

We can now weight these probabilities with the initial density matrix elements to get the full probability of crossing, $P_{cross} = p_E P_{cross}^E + p_O P_{cross}^O$.  
Combining results, we find the entanglement genesis probability is given by
\be
P_{EG} = (\rho_{11} + \rho_{22}) e^{2 r_2} + (\rho_{33} + \rho_{44}) = 
2 \max [\rho_{33},\rho_{44}],
\ee  
where we used the threshold (\ref{rs}).  Analogously, the sudden death probability is given by
\bea
P_{SD} &=& (\rho_{11} + \rho_{22}) + (\rho_{33} + \rho_{44})e^{-2 r_2}, \nonumber \\
&=&  (\rho_{11} + \rho_{22})\frac{2 \max [\rho_{33},\rho_{44}]}{|\rho_{33}-\rho_{44}|}.
\eea
Accounting for normalization recovers Eq.~(\ref{sdprob}).

We can renormalize \eqref{fptprob} with \eqref{crossingprob} to get the probability distributions $P_{fpt|C}^i$ of crossing as a function of time, conditioned on the fact that they will cross.  This will allow us to determine the first passage time statistics.  These conditional probabilities are
\be
P_{fpt|C}^i(\tau) = \frac{|r_2|}{\sqrt{4 \pi D \tau^3}} \exp \Big( - \frac{(|r_2| - |v_i|\tau)^2}{4 D \tau} \Big).
\label{crossprob}
\ee
Using this probability density, the average first passage time is given by  $\la \tau_i \ra = \int_0^\infty d\tau \, \tau P_{fpt|C}^i =  |r_2|/|v_i|$.  When this average time is weighted by the initial density matrix elements, the total average time until the border crossing is 
\be
 \la \tau \ra = p_E \tau_E + p_O \tau_O= |r_2|.
\ee
 Restoring physical time units gives the result (\ref{avgenttime}).

\acknowledgments
We thank Joe Eberly and Ting Yu for enlightening discussions.  This work was supported by the University of Rochester.

\end{document}